\documentclass[a4paper,11pt]{article}
\usepackage{pos}
\usepackage{natbib}

\title{Two-loop corrections to the Higgs trilinear coupling in classically scale-invariant theories}
\ShortTitle{Two-loop corrections to the Higgs trilinear coupling in CSI theories}

\author*[a]{Johannes Braathen}
\author[b]{Shinya Kanemura}
\author[b]{Makoto Shimoda}

\affiliation[a]{Deutsches Elektronen-Synchrotron DESY,\\ Notkestra\ss{}e 85, 22607 Hamburg, Germany.}

\affiliation[b]{Department of Physics, Osaka University,\\
Toyonaka, Osaka 560-0043, Japan.\vspace{.2cm}}

\emailAdd{johannes.braathen@desy.de}
\emailAdd{kanemu@het.phys.sci.osaka-u.ac.jp}
\emailAdd{m\_shimoda@het.phys.sci.osaka-u.ac.jp}

\abstract{
The Higgs trilinear coupling is a crucial tool to investigate the structure of the Higgs sector and the nature of the electroweak phase transition, and to search for indirect signs of New Physics. Classical scale invariance (CSI) is an attractive concept for BSM model building, explaining the apparent alignment of the Higgs sector and potentially relating to the hierarchy problem. A particularly interesting feature of CSI theories is that, at one loop, they universally predict the Higgs trilinear coupling to deviate by 67\% from the SM prediction at tree level.
This result is however modified at two loops, and we present here results from the first explicit computation of two-loop corrections to the Higgs trilinear coupling in classically scale-invariant BSM models. Taking as example a CSI variant of the Two-Higgs-Doublet Model, we show that the inclusion of two-loop effects allows distinguishing different scenarios with CSI, even though the requirement of correctly reproducing the mass of the Higgs boson, as well as unitarity, severely restrict the possible values of the Higgs trilinear coupling.
}

\FullConference{%
  *** The European Physical Society Conference on High Energy Physics (EPS-HEP2021), ***\\
  *** 26-30 July 2021 ***\\
  *** Online conference, jointly organized by Universität Hamburg and the research center DESY ***
}

\begin{document}
\begin{flushright}
DESY-21-159\\
OU-HET-1110
\end{flushright}
\maketitle

\section{Introduction}

The Higgs trilinear coupling $\lambda_{hhh}$ offers a unique opportunity to probe the Higgs sector and search for signs of Beyond-the-Standard-Model (BSM) physics. Firstly, it directly relates to the shape of the Higgs potential, of which little is currently known. Indeed, only the location of the electroweak (EW) minimum -- obtained from the Higgs vacuum expactation value (VEV) -- and the curvature of the potential around this minimum -- given by the Higgs mass -- are known. In particular, the behaviour of the potential away from the EW minimum remains to be determined -- and it depends greatly on $\lambda_{hhh}$. In turn, the strength of the EW phase transition (EWPT) also depends on the value of $\lambda_{hhh}$: for instance in order for the EWPT to be of strong first order (a necessary condition $e.g.$ for successful EW baryogenesis), $\lambda_{hhh}$ must deviate by at least 20\% from its SM prediction~\cite{Grojean:2004xa,Kanemura:2004ch}. Moreover, because its experimental determination is currently not very precise but will be drastically improved in the future (for a review see Ref.~\cite{deBlas:2019rxi}), the Higgs trilinear coupling is an ideal target to search for large BSM deviations~\cite{Kanemura:2004mg}.

Theories with classical scale invariance (CSI), in which all Lagrangian mass-dimensionful are forbidden at tree level, provide a good example of this. In these models, a flat direction must exist in the tree-level potential, along which the EW gauge symmetry is broken radiatively \`{a} la Coleman-Weinberg~\cite{Coleman:1973jx,Gildener:1976ih}. This flat direction corresponds to the 125-GeV Higgs boson and is automatically aligned at tree level~\cite{Lane:2018ycs} (see also Ref.~\cite{Eichten:2021qbm} for a discussion of how this is modified at loop level). Furthermore, BSM states cannot be decoupled as all Lagrangian mass terms are forbidden -- thereby making CSI theories perfect examples of scenarios with \emph{alignment without decoupling}~\cite{Gunion:2002zf}.
Another distinctive property of CSI models is that, at one loop, $\lambda_{hhh}$ is \emph{universally} predicted to deviate by 67\% from the tree-level SM value of $\lambda_{hhh}$~\cite{Hashino:2015nxa}. In Ref.~\cite{Braathen:2020vwo}, we found that this universality is lost once two-loop effects are included. We summarise in these proceedings our computation of the dominant two-loop corrections to $\lambda_{hhh}$, and we present some examples of our numerical results in a CSI variant of the Two-Higgs-Doublet Model (2HDM).

\section{The Higgs trilinear coupling at one and two loops}

We begin by reviewing briefly the calculation in CSI models of $\lambda_{hhh}$ -- defined in terms of the effective potential $V_\text{eff}$ as $\lambda_{hhh}\equiv\frac{\partial^3V_\text{eff}}{\partial h^3}\big|_\text{min}$ -- and we recall the most important results at one and two loops. The special results for $\lambda_{hhh}$ stem from the particular form of field-dependent masses in CSI models: in the absence of any BSM mass-dimensionful term, the field-dependent masses of a state $i$ -- no matter its nature -- can be written as $m_i^2(h) =m_i^2(1+h/v)^2$, where $m_i$ is the corresponding field-independent mass and $v$ is the Higgs VEV. This implies that along the Higgs direction in field space, the one-loop effective potential takes the very simple form
\begin{align}
V_\text{eff}(h)=A\cdot (v+h)^4+B \cdot(v+h)^4\log\frac{(v+h)^2}{Q^2}\,,
\end{align}
where $Q$ is the renormalisation scale, and $A$, $B$ are functions of the scalar, fermion, and gauge-boson mass matrices of the model considered -- their expressions can be found $e.g.$ in Ref.~\cite{Hashino:2015nxa}. When performing an effective-potential computation, we can express first $A$ in terms of $B$ and $v$ with the tadpole condition, and next we can eliminate $B$ in favour of the Higgs effective-potential mass $[M_h^2]_{V_\text{eff}}$ and $v$. In other words, the one-loop potential is entirely fixed in terms of the Higgs mass and VEV, which are both known experimentally. As a consequence, one finds that at one loop the Higgs trilinear coupling is \emph{universally} predicted in CSI models to be~\cite{Hashino:2015nxa} 
\begin{align}
\label{EQ:lambdahhh_1L}
\lambda_{hhh}=\frac{5[M_h^2]_{V_\text{eff}}}{v}=\frac{5}{3}(\lambda_{hhh}^\text{SM})^\text{tree}\,,
\end{align}
$(\lambda_{hhh}^\text{SM})^\text{tree}\simeq 191\text{ GeV}$ being the tree-level SM prediction. We emphasise that this result is completely independent of the particle content of the CSI model. 
This picture is however altered once one includes two-loop corrections to $V_\text{eff}$ and $\lambda_{hhh}$. Indeed, new types of terms involving squared logarithms appear in $V_\text{eff}$ at two loops (see $e.g.$ Ref.~\cite{Ford:1992pn}) and the two-loop potential takes the form
\begin{align}
V_\text{eff}(h)=A\cdot (v+h)^4+B \cdot(v+h)^4\log\frac{(v+h)^2}{Q^2}+C \cdot(v+h)^4\log^2\frac{(v+h)^2}{Q^2}\,.
\end{align}
$A$ and $B$ receive both one- and two-loop contributions but $C$ is a purely two-loop quantity. 
Like at one loop, $A$ and $B$ can be eliminated, using respectively the tadpole equation and the Higgs effective-potential mass, however $C$ remains and one finds that at two loops
\begin{align}
\label{EQ:lambdahhh_2L}
\lambda_{hhh}=\frac{5[M_h^2]_{V_\text{eff}}}{v}+32C v\,.
\end{align}
Because the effective potential is model-dependent, it follows that $C$, and hence also $\lambda_{hhh}$, are also model-dependent -- $i.e.$ the universality of $\lambda_{hhh}$ found at one loop is \emph{lost} once two-loop effects are taken into account. We refer the interested reader to Ref.~\cite{Braathen:2020vwo} for technical details of our derivations and complete expressions for our results. We only mention here that general results for two-loop contributions to $V_\text{eff}$, written in terms of $\overline{\text{MS}}$-renormalised parameters, can be found $e.g.$ in Ref.~\cite{Martin:2001vx}. These can be used to extract the $\log^2$ coefficient $C$, from which $\lambda_{hhh}$ is straightforwardly obtained (in the $\overline{\text{MS}}$ scheme) using equation~(\ref{EQ:lambdahhh_2L}). Additionally, we included the necessary finite counterterms to translate our expressions to the OS scheme.  

\section{Numerical analysis}

\begin{figure}
	\centering
	\includegraphics[width=.48\textwidth]{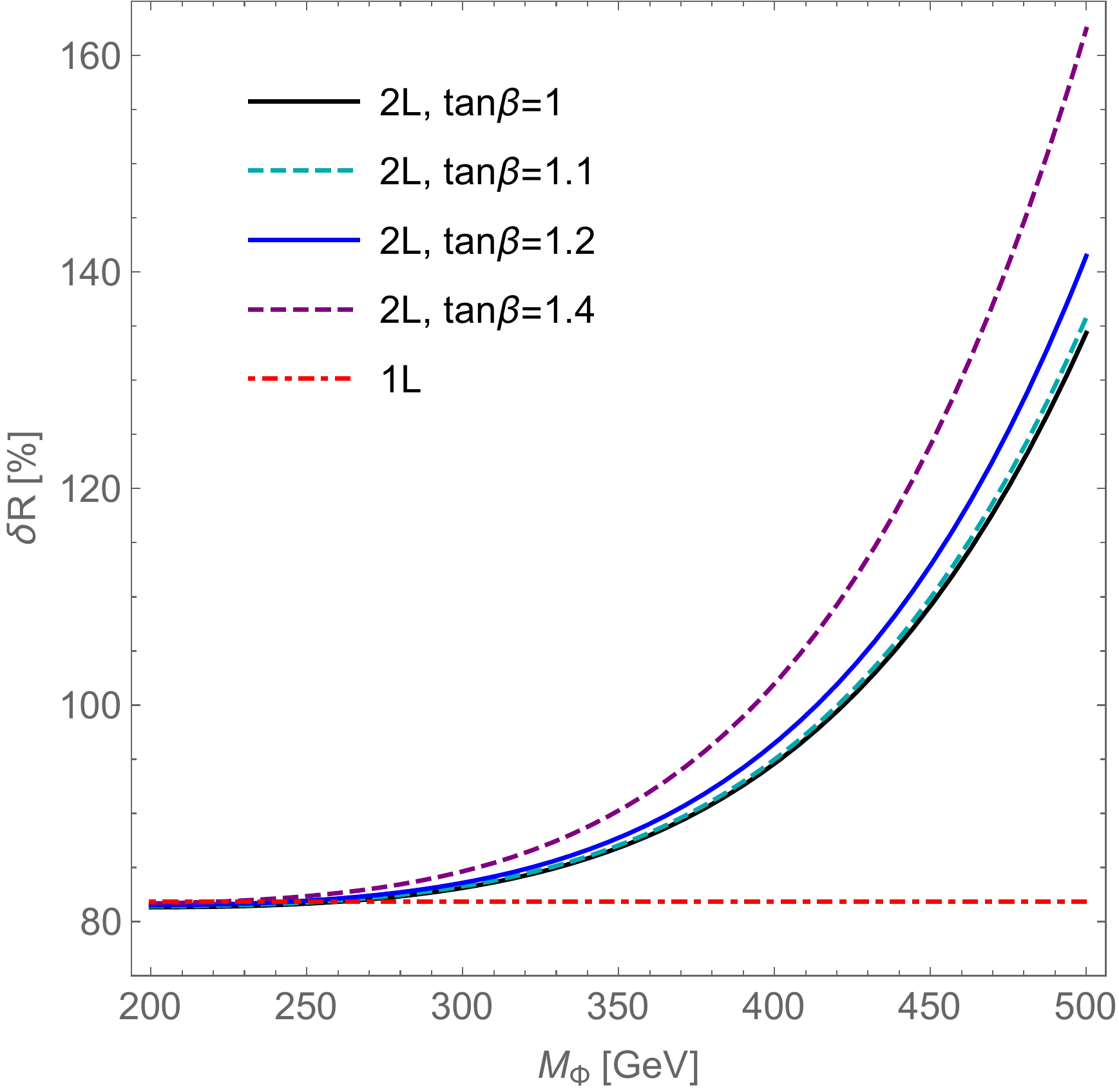}
	\includegraphics[width=.48\textwidth]{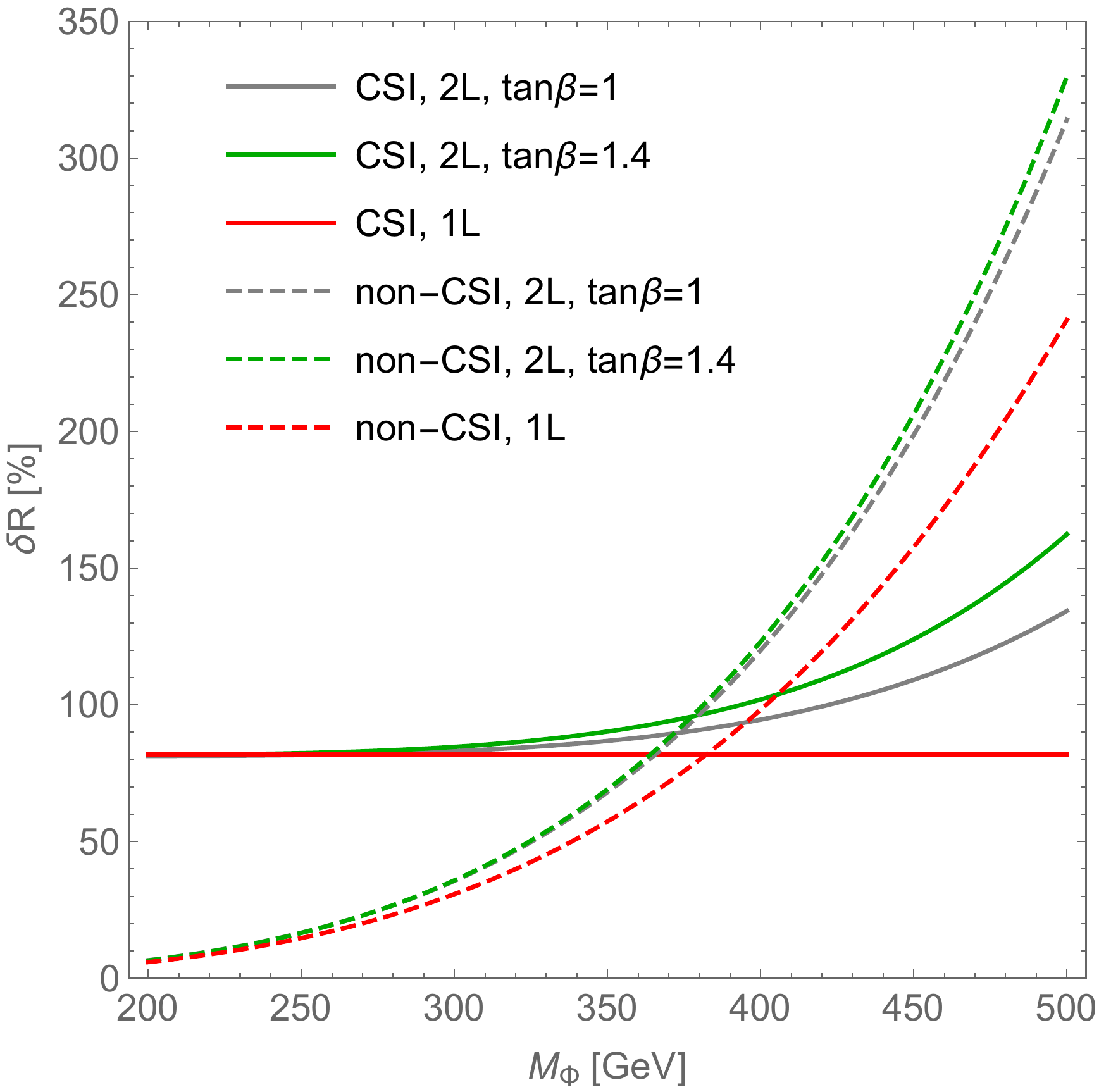}
	\caption{BSM deviation $\delta R$ in the Higgs trilinear coupling, as a function of the degenerate mass $M_\Phi$ of the BSM scalars. \emph{Left}: Deviation in the CSI-2HDM at one loop (red) and at two loops for different values of $\tan\beta$. \emph{Right}: Comparison between the deviations in the CSI-2HDM (solid curves) and in an aligned scenario of the usual 2HDM (dashed curves). One-loop results are in red while grey and green curves correspond to two-loop values respectively for $\tan\beta=1$ and $\tan\beta=1.4$. }
	\label{FIG:var_MPhi_tanb}
\end{figure}

As a concrete setting to present numerical results, we consider a CSI variant~\cite{Lee:2012jn} of the (CP-conserving) 2HDM. Like the normal 2HDM, the CSI-2HDM contains three additional Higgs bosons -- $H$ (CP-even), $A$ (CP-odd), and $H^\pm$ (charged) -- however it differs from the usual model by absence of mass terms in the scalar potential and the automatic alignement of its Higgs sector at tree level. More details on the model and our conventions can be found in Ref.~\cite{Braathen:2020vwo}. We present in the following results for the BSM deviation of the trilinear coupling computed in the CSI-2HDM with respect to the SM, which we define as $\delta R\equiv \lambda_{hhh}^\text{CSI-2HDM}/\lambda_{hhh}^\text{SM}-1$ (the values of $\lambda_{hhh}$ being computed at the \emph{same order} in both models). The left side of figure~\ref{FIG:var_MPhi_tanb} shows the deviation $\delta R$ as a function of the degenerate mass $M_\Phi$ of the BSM scalars in the CSI-2HDM ($M_\Phi=M_H=M_A=M_{H^\pm}$) at one loop (red line) and at two loops for different values of $\tan\beta$ (black, cyan, blue, and purple curves) -- $\tan\beta$ being the ratio of the VEVs of the neutral components of the two Higgs doublets. While the one-loop result is constant\footnote{We have a deviation of $82\%$, and not $67\%$ as shown in eq.~(\ref{EQ:lambdahhh_1L}), because we compare here the one-loop result in the CSI-2HDM, with the \emph{one-loop} (rather than tree-level) SM prediction. } as explained in the previous section, the inclusion of two-loop effects in $\delta R$ introduces a significant dependence on both $M_\Phi$ and $\tan\beta$, and hence allows distinguishing different parameter points of CSI-2HDM using the value of $\lambda_{hhh}$. Furthermore, the two-loop corrections to $\lambda_{hhh}$ result in an additional positive -- and potentially large for increasing $M_\Phi$ -- shift in $\delta R$, meaning that $\lambda_{hhh}$ is more easily accessible in experiments than what would be expected from the one-loop result. 

In the right side of figure~\ref{FIG:var_MPhi_tanb}, we compare the behaviours of $\delta R$ as a function of $M_\Phi$ between the CSI-2HDM and an aligned scenario of the usual 2HDM (in the maximal non-decoupling limit $M=0$~\cite{Kanemura:2004mg}) -- where for the latter we employ expressions from Refs.~\cite{Braathen:2019}. Due to the drastically different behaviours at one loop -- an 82\% deviation in the CSI case, compared to a growing deviation proportional to $M_\Phi^4$ in the usual 2HDM -- the total deviations at two loops are distinguishable for most of the considered range of $M_\Phi$, although the two-loop corrections to $\lambda_{hhh}$ behave similarly in the two scenarios, scaling like $M_\Phi^6$. For low BSM masses, the largest effects are found in the CSI case while, conversely, for large masses the non-CSI scenario exhibits the most significant deviations (driven by the growth of the non-decoupling effects at one loop).

\begin{figure}
	\centering
	\includegraphics[width=.7\textwidth]{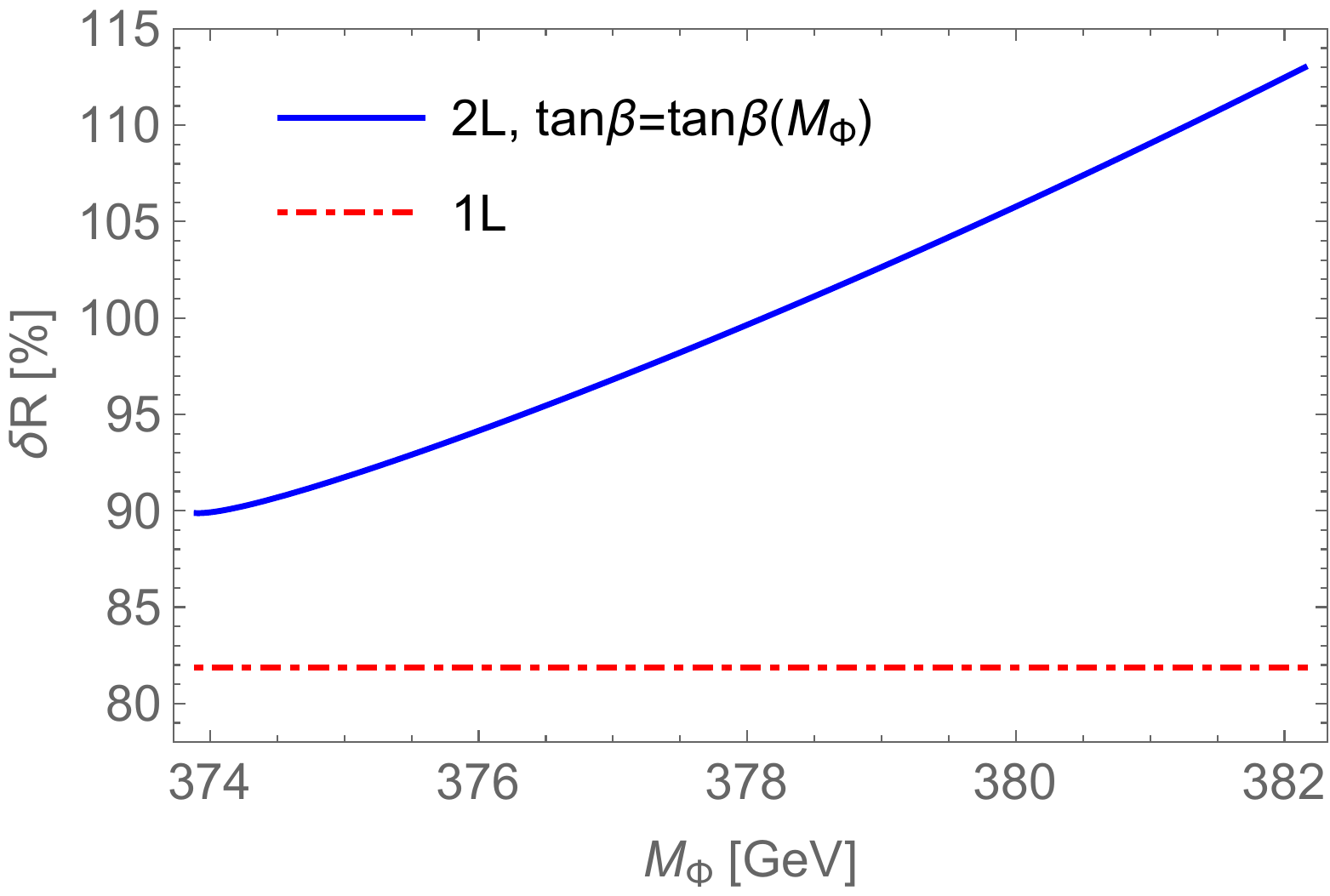}
	\caption{BSM deviation $\delta R$ as a function of the degenerate mass $M_\Phi$ of the BSM scalars. For the two-loop result (blue), $\tan\beta$ is determined in terms of $M_\Phi$ from the requirement of reproducing $M_h=125\text{ GeV}$. }
	\label{FIG:constrained}
\end{figure}

In both plots of figure~\ref{FIG:var_MPhi_tanb}, we have verified that perturbative unitarity~\cite{Kanemura:1993hm} is maintained and also that the EW minimum is the true minimum of the effective potential. A further important constraint that must be considered as well is the requirement of correctly reproducing the 125-GeV mass of the SM-like Higgs boson: indeed as the Higgs boson corresponds to the flat direction of the tree-level potential, its mass must be generated entirely at loop level. This yields a relation between the known SM inputs and the BSM parameters, from which one of the latter can be extracted; in the CSI-2HDM, $\tan\beta$ can thence be obtained as a function of $M_\Phi$. In figure~\ref{FIG:constrained}, we show the possible values of $\delta R$, again as a function of $M_\Phi$, once $\tan\beta$ is fixed by the Higgs-mass constraint. This additional requirement, combined with perturbative unitarity, severely limits the allowed range of the BSM scalar mass $M_\Phi$, which must now remain between 374 and 382 GeV. At two loops, $\lambda_{hhh}$ is found to deviate from the SM prediction by 90\% to 113\% -- $i.e.$ approximately 10\% to 33\% more than at one loop. Interestingly, if one returns to the right pane of figure~\ref{FIG:var_MPhi_tanb}, one notices that this range of $M_\Phi$ and $\delta R$ corresponds to where the curves from the CSI and non-CSI variants of the 2HDM overlap, meaning a measurement of $\lambda_{hhh}$ may not suffice by itself to ascertain whether a BSM scenario exhibits CSI or not. Finally, the parameter points presented in figure~\ref{FIG:constrained} have been checked against limits from experimental searches with \texttt{HiggsBounds}~\cite{HiggsBounds} (the necessary inputs were produced by a \texttt{SPheno}~\cite{SPheno} based spectrum generator for the CSI-2HDM created using \texttt{SARAH}~\cite{SARAH}). 
\vspace{-.25cm} 
 
\section{Conclusions}
\vspace{-.25cm}
In these proceedings, we have summarised our results from Ref.~\cite{Braathen:2020vwo}, in which the impact of the leading two-loop corrections to the Higgs trilinear coupling in models with CSI was analysed. 
Our most important finding is that the inclusion of two-loop contributions in $\lambda_{hhh}$ allows distinguishing different parameter points of a given CSI scenario, because the universality of the prediction for $\lambda_{hhh}$, found at one loop, is lifted at two loops. However, we also pointed out how different theory constraints -- in particular perturbative unitarity and the need to generate the correct mass of 125 GeV for the Higgs boson -- severely limit the allowed range of BSM parameters, and in turn the possible values of $\delta R$. Once these constraints are included, the two-loop prediction for $\lambda_{hhh}$ in the CSI-2HDM is found to deviate from the SM value by 90-113\% -- $i.e.$ 10-33\% more than at one loop. An adverse consequence of this is that a measurement of $\lambda_{hhh}$ may not suffice by itself to distinguish CSI or non-CSI versions of a given model -- although this could be achieved by using the synergy of such a measurement with either collider or gravitational-wave searches (see $e.g.$ Ref.~\cite{Hashino:2016rvx}). Finally, while we illustrated the present discussion with numerical investigations for a CSI variant of the 2HDM, our findings apply to more broadly to CSI models -- in particular in Ref.~\cite{Braathen:2020vwo} similar results were also obtained in $N$-scalar models. 

\subsection*{Acknowledgments}
This work is supported by JSPS, Grant-in-Aid for Scientific Research, No. 16H06492, 18F18022, 18F18321 and 20H00160. This work is also partly supported by the Deutsche Forschungsgemeinschaft (DFG, German Research Foundation) under Germany’s Excellence Strategy – EXC 2121 “Quantum Universe” – 390833306.

\end{document}